\documentstyle[preprint,aps]{revtex}
\begin{document}
\title{Scattering off an SO(10) cosmic string}
\author{Anne-Christine DAVIS\thanks{and King's College, Cambridge}
and Rachel JEANNEROT,\\
        {\normalsize{ Department of Applied Mathematics and
Theoretical Physics,}}\\
	{\normalsize{ Cambridge University,}}\\
        {\normalsize{ Silver Street, Cambridge, CB3 9EW, UK}}\\
	{\normalsize{ and}}\\
	{\normalsize{ Isaac Newton Institute for Mathematical
Sciences,}}\\
	{\normalsize{ Cambridge University,}}\\
        {\normalsize{ 20 Clarkson Road, Cambridge, CB3 0EH, UK.}}}
\maketitle

\begin{abstract}
The scattering of fermions from the abelian string
arising during the phase transition $SO(10)
\rightarrow SU(5) \times Z_2$ induced by the Higgs in
the 126 representation is studied. Elastic
cross-sections and baryon number violating
cross-sections due to the coupling to gauge fields in
the core of the string are computed by both a first
quantised method and a perturbative second quantised
method. The elastic cross-sections are found to be
Aharonov-Bohm type. However, there is a marked
asymmetry between the scattering cross-sections for
left and right handed fields. The catalysis
cross-sections are small, depending on the grand
unified scale. If cosmic strings were observed our
results could help tie down the underlying gauge
group.
\end{abstract}

\section{Introduction}

Modern particle physics and the hot big-bang model
suggest that the universe underwent a series of phase
transitions at early times at which the underlying
symmetry changed. At such phase transitions
topological defects~\cite{Shellard94} could be
formed. Such topological defects, in particular cosmic
strings, would still be around today and provide a
window into the physics of the early universe. In
particular, cosmic strings arising from a grand
unified phase transition are good candidates for the
generation of density perturbations in the early
universe which lead to the formation of large scale
structure~\cite{Robert}. They could also give rise to
the observed anisotropy in the microwave background
radiation~\cite{Joao}.

Cosmic strings also have interesting microphysical
properties. Like monopoles~\cite{Rubakov} they can
catalyse baryon violating
processes~\cite{Branden88,Perkins91}. This is because
the full grand unified symmetry is restored in the
core of the string, and hence grand unified, baryon
violating processes are
unsuppressed. In~\cite{Perkins91} it was shown that
the cosmic string catalysis cross-section could be a
strong interaction cross-section, independent of the
grand unified scale, depending on the flux on the
string. Unlike the case of monopoles, where there is a
Dirac quantisation condition, the string cross-section
is highly sensitive to the flux, and is a purely
quantum phenomena. Defect catalysis is potentially
important. It has already been used to bound the
monopole flux~\cite{Warren}, and could erase a
primordial baryon asymmetry~\cite{Branden89}. It is,
thus, important to calculate the string catalysis
cross-section in a realistic grand unified
theory. In~\cite{Perkins91} a toy model based on a
$U(1)$ theory was used. In a grand unified theory the
string flux is given by the gauge group, and cannot be
tuned.

A cosmic string is essentially a flux tube. Hence the
elastic cross-section~\cite{Alf89} is just an
Aharonov-Bohm cross-section~\cite{Aharonov}, depending
on the string flux. This gives the dominant energy
loss in a friction dominated
universe~\cite{Adrian}. Since the string flux is fixed
for any given particle species it is important to
check that the Aharonov-Bohm cross-section persists in
a realistic grand unified theory.

In this paper we calculate the elastic and inelastic
cross-sections for cosmic strings arising from an
$SO(10)$ grand unified theory~\cite{G74}. Cosmic
strings arise in the breaking scheme~\cite{Kibble82}
$SO(10) \rightarrow SU(5) \times Z_2$ where the
breaking is due to the $126$ representation of the
Higgs field, the self-dual anti-symmetric 5-index
tensor of SO(10). These stable strings survive the
subsequent transitions to $SU(3) \times SU(2) \times
U(1) \times Z_2$ ~\cite{Kibble82}. They have been
studied elsewhere ~\cite{Aryal87}.

Now the SO(10) symmetry is restored inside the string
core, and therefore there are baryon number violation
processes mediated by the gauge fields $X$, $Y$, $X'$,
$Y'$ and $X_s$ of SO(10). We therefore expect a
non-zero inelastic cross-section which we will
determine. This cross-section should be running from a
small cross-section $O(\eta^{-1})$, where $\eta$ is
the grand unified scale $\sim 10^{15}$ GeV to a much
larger cross-section of the order of the strong
interaction.

The plan of this paper is as follow: In
section~\ref{sec-string} we define an SO(10) string
model. We give 'top-hat' forms for the Higgs and gauge
fields forming the string, since the 'top-hat' core
model doesn't affect the cross-sections of interest
structure of the string core, we introduce the baryon
number violating gauge fields of SO(10) present in the
core of the string.

In section~\ref{sec-cs} we review the method used to
calculate the scattering cross-sections. There are two
different approaches. A fundamental quantum mechanical
one and a perturbative second quantised method
calculating the geometrical cross-section, i.e. the
scattering cross-section for free fermionic
fields. The catalysis cross-section is then enhanced
by an amplification factor to the power of four.

In section~\ref{sec-motion} we derive the equations of
motion. In order to simplify the calculations and to
get a fuller result, we also consider a 'top-hat' core
model for the gauge fields mediating quark to lepton
transitions.

In section~\ref{sec-ext} and section~\ref{sec-int} we
calculate the solutions to the equations of motion
outside and inside the string core respectively, and
in section~\ref{sec-match} we match our solutions at
the string radius. In section~\ref{sec-ampl} we
calculate the scattering amplitude for incoming plane
waves of linear combinations of the quark and electron
fields.

We use these results in section~\ref{sec-elast} and
section~\ref{sec-inelast} in order to calculate the
scattering cross-sections of incoming beams of pure
single fermion fields. In section~\ref{sec-elast} we
calculate the elastic cross-sections. And in
section~\ref{sec-concl} we calculate the baryon number
violation cross-sections.

In section~\ref{sec-second} we derive the catalysis
cross-section using the second quantised method of
ref.~\cite{Branden88,Perkins91}. The second-quantised
cross-sections are found to agree with the first
quantised cross-section of section~\ref{sec-inelast}.

There are 4 appendices. Appendix~\ref{sec-so10} gives
a brief review on SO(10) theory, and gives an explicit
notation used everywhere in this
paper. Appendices~\ref{sec-extap} and~\ref{sec-intap}
contain the technical details of the external and
internal solutions calculations. Finally,
Appendix~\ref{sec-matchap} is a discussion of the
matching conditions at the core radius.

\section{An SO(10) string}
\label{sec-string}

In the appendix A, we give a brief review of SO(10)
theory. With that notation, the lagrangian is,
\begin{equation}
L = {1\over 4} F_{\mu\nu} F^{\mu\nu} +
(D_\mu\Phi_{126})^{\dagger} (D^\mu\Phi_{126})  -
V(\Phi) + L_F \label{eq:lagrangian}
\end{equation}
where $F_{\mu\nu} = -i\, F_{\mu\nu}^a \tau_a$, $\tau_a
\, \, a = 1,...,45$  are the 45 generators of
SO(10). $\Phi_{126}$ is the Higgs 126,  the self-dual
anti-symmetric 5-index tensor of SO(10). $L_F$ is the
fermionic part of the lagrangian. In the covariant
derivative $D_\mu = \partial_\mu + i e A_\mu$, $A_\mu =
A_\mu^a \tau_a$ where $A_\mu^a$ a = 1,...,45 are 45
gauge fields of SO(10).

We assume that the universe undergoes the following
breaking scheme,
$$
SO(10) \stackrel{<\phi_{126}>}{\rightarrow} SU(5)
\times Z_2 \stackrel{<\phi_{45}>}{\rightarrow} SU(3)
\times SU(2) \times U(1) \times Z_2
\stackrel{<\phi_{10}>}{\rightarrow} SU(3) \times
U(1)_Q \times Z_2
$$
giving vacuum expectation values to the components of
the 10 which correspond to the usual Higgs
doublet. The decomposition of the 126 representation
under $SU(5) \times U(1)$ is given by,
\begin{equation}
126 = 1_{10} + ... \:.
\end{equation}
The first transition is achieved by giving vacuum
expectation value to the component of the 126 in the
$1_{10}$ direction. The first homotopy group $\pi_1
(SO(10) / SU(5) \times Z_2)$ is $Z_2$, and therefore
$Z_2$ strings are formed.  In terms of SU(5), the 45
generators of SO(10) can be decomposed as follows,
\begin{equation}
45 = 24 + 1 + 10 + \bar{10} \:
\end{equation}
 From the 45 generators of SO(10), 24 belong to SU(5),
1 generator corresponds to the $U(1)'$ symmetry in
SO(10) not embedded in SU(5) and there are 20
remaining ones. Therefore the breaking of SO(10) to
$SU(5) \times Z_2$ induces the creation of two types
of strings. An Abelian one, corresponding to the
$U(1)'$ symmetry, and an non abelian one made with
linear combinations of the 20 remaining generators. In
this paper we are interested in the abelian strings
since the non abelian version are Alice strings, and
would result in global quantum number being
ill-defined, and hence unobservable \cite{Bucher}. We
note that there is a wide range of parameters where
the non abelian strings have lower energy
\cite{Aryal87}. However, since the abelian string is
topologically stable, there is a final probability
that it could be formed by the Kibble mechanism
\cite{Kib76}.

If we call $\tau_{str}$ the generator of the abelian
string, $\tau_{str}$ will be given by the diagonal
generators of SO(10) not lying in $SU(5)$ that is,
\begin{equation}
\tau_{str} = {1\over 5} \, (M_{12} + M_{34} + M_{56} + M_{78} + M_{9\, 10})
\end{equation}
where $M_{ij} : i,j = 1...10$ are the 45 SO(10)
generators defined in appendix A in terms of the
generalised gamma matrices. Numerically, this gives,
\begin{equation}
\tau_{str} = diag({1\over 2}\, , {1\over 10}\, ,
{1\over 10}\, , {1\over 10}\, , {1\over 10}\, ,
{1\over 10}\, , {1\over 10}\, , {-3\over 10}\, ,
{1\over 10}\, , {1\over 10}\, , {1\over 10}\, ,
{-3\over 10}\, , {1\over 10}\, , {-3\over 10}\, ,
{-3\over 10}\, ,  {-3\over 10}) \label{eq:tauall} \:
{}.
\end{equation}
The results of Perkins et Al. \cite{Perkins91} find
that the greatest enhancement of the cross-section is
for fermionic charges close to integer values. Thus,
from equation (\ref{eq:tauall}), we expect no great
enhancement; the most being due to the right-handed
neutrino.

We are going to model our string as is usually done
for an abelian U(1) string. That is, we take the
string along the z axis, resulting in the Higgs
$\Phi_{126}$ and the gauge fields $A_\mu$ of the
string to be independent of the z coordinate,
depending only on the polar coordinates $(r,
\theta)$. Here $A_\mu$ is the gauge field of the
string, obtained from the product $A_\mu = A_{\mu ,
str} \tau_{str}$. The solution for the abelian string
can be written as,
\begin{eqnarray}
\Phi_{126} &=& f(r) \, e^{i \tau_{str} \theta} \,
\Phi_0 = f(r) \, e^{i \theta} \, \Phi_0 \\
A_\theta &=& - {g(r)\over er} \tau_{str}\nonumber\\
A_r&=& A_z=0 \label{eq:gauge}
\end{eqnarray}
where $\Phi_0$ is the vacuum expectation value of the
Higgs 126 in the $1_{10}$ direction. The functions
$f(r)$ and $g(r)$ describing the behaviour the Higgs
and gauge fields forming the string are given by
\begin{eqnarray}
\left. \begin{array}{lllllll}

f(r) &=& \left\{ \begin{array} {cr}
                  {\eta } & r \geq R\\
                  {\eta }  ({r\over R}) & r < R
                  \end{array}
          \right.
& , & g(r) &=& \left\{ \begin{array} {cr}
                  1 & r\geq R\\
                  ({r\over R})^2 & r < R
                  \end{array}
          \right.
\end{array}
 \right.
\end{eqnarray}
where R is the radius of the string. $R \sim
\eta^{-1}$, where $\eta$ is the grand unified scale,
assumed to be $\eta \sim 10^{15} GeV$. In order to
simplify the calculations and to get a fuller result
we use the top-hat core model, since it has been shown
not to affect the cross-sections of interest. The
top-hat core model assumes that the Higgs and gauge
fields forming the string are zero inside the string
core. Hence, $f(r)$ and $g(r)$ are now given by,
\begin{eqnarray}
\left. \begin{array}{lllllll}
f(r) &=& \left\{ \begin{array} {cr}
                  {\eta } & r \geq R\\
                   0 & r < R
                  \end{array}
          \right.
& , & g(r) &=& \left\{ \begin{array} {cr}
                  1 & r\geq R\\
                  0 & r < R
                  \end{array}
          \right.  \: .
 \end{array}
 \right. \label{eq:tophat}
\end{eqnarray}

The full $SO(10)$ symmetry is restored in the core of
the string. $SO(10)$ contains 30 gauge bosons
leading to baryon decay. These are the bosons $X$ and
$Y$, and their conjugates, of SU(5) plus 18 other gauge
bosons usual called $X'$, $Y'$ and $X_s$, and their
conjugates. Therefore inside the
string core, there are quark to lepton transitions
mediated by the gauge bosons $X$, $X'$, $Y$, $Y'$ and
$X_s$ and we expect the string to catalyse baryon
number violating processes in the early universe.

The $X$, $X'$, $Y$, $Y'$ and $X_s$ gauge bosons are
associated with non diagonal generators of SO(10). For
the electron family, the relevant part of the
lagrangian is given by,
\begin{equation}
L_x = \bar{\Psi}_{16} \, ( i e \gamma^\mu (X_\mu \tau^X + X'_\mu \tau^{X'}
 + Y_\mu \tau^Y  + Y'_\mu \tau^{Y'} + {X_s}_\mu
\tau^{X_s})) \, \Psi_{16} \label{eq:baryon}
\end{equation}
where $\tau^X$, $\tau^{X'}$, $\tau^Y$, and $\tau^{Y'}$
and $\tau^{X_s}$ are the non diagonal generators of
SO(10) assosiated with the  $X$, $X'$, $Y$, $Y'$ and
$X_s$ gauge bosons respectively.

Expending equation (\ref{eq:baryon}) gives ~\cite{Marie},
\begin{eqnarray}
L_x &=& {g\over\sqrt{2}} X^{\alpha }_\mu [-\epsilon_{\alpha \beta \gamma }
\bar{u}^{c\gamma }_L \gamma^\mu u^\beta_L + \bar{d}_{L \alpha}
\gamma^\mu e^+_L +  \bar{d}_{R \alpha} \gamma^\mu e^+_R
]\nonumber \\
&& + {g\over\sqrt{2} } Y^{\alpha }_\mu [-\epsilon_{\alpha \beta \gamma }
\bar{u}^{c\gamma }_L \gamma^\mu d^\beta_L - \bar{d}_{R \alpha}
\gamma^\mu \nu_e^c - \bar{u}_{L\alpha } \gamma^\mu e^+_L ]\nonumber
\\
&& + {g\over\sqrt{2} } X^{\alpha '}_\mu  [-\epsilon_{\alpha
\beta \gamma } \bar{d}^{c\gamma }_L \gamma^\mu d^\beta_L -
\bar{u}_{R\alpha} \gamma^\mu \nu_R^c - \bar{u}_{L\alpha} \gamma^\mu
\nu_L^c ] \nonumber\\
&&  + {g\over\sqrt{2} } Y^{\alpha '}_\mu [\epsilon_{\alpha \beta \gamma }
\bar{d}^{c\gamma }_L \gamma^\mu u^\beta_L - \bar{u}_{R\alpha }
\gamma^\mu e_R^+ - \bar{d}_{L\alpha } \gamma^\mu \nu_L^c ]
\nonumber\\
&& + {g\over\sqrt{2} } X^{\alpha }_{s \mu} [\bar{d}_{L \alpha }
\gamma^\mu e_L^- + \bar{d}_{R\alpha } \gamma^\mu e_R^- +
\bar{u}_{L\alpha } \gamma^\mu \nu_L + \bar{u}_{R \alpha } \gamma^\mu
\nu_R ]  \label{eq:lagi}
\end{eqnarray}
where $\alpha$, $\beta$ and $\gamma$ are colour indices. The
$X_s$ does not contribute to nucleon decay except by
mixing with the $X'$ because there is no vertex
$qqX_s$. We consider baryon violating processes
mediated by the gauge fields $X$, $X'$, $Y$ and $Y'$
of SO(10). In previous papers ~\cite{Perkins91,Alf89},
baryon number violating processes resulting from the
coupling to scalar condensates in the string core have
been considered. In our SO(10) model we do not have
such a coupling.

\section{Scattering of Fermions from the Abelian String}
\subsection{The scattering cross-section}
\label{sec-cs}
Here, we will briefly review the two methods used to
calculate the scattering cross-section. The first is a
quantum mechanical treatment. From the fermionic
lagrangian $L_F$, we derive the equations of motion
inside and outside the string core. We then find
solutions to the equations of motion inside and
outside the string core and we match our solutions at
the string core. Considering incoming plane waves of
pure quarks, we then calculate the scattering
amplitude. The matching conditions together with the
scattering amplitude enable us to calculate the
elastic and inelastic scattering cross-sections. The
second method is a quantised one, where one calculates
the geometrical cross-section $({d\sigma \over d
\Omega})_{geom}$, i.e. using free fermions spinors
$\psi_{free}$. The catalysis cross-section is enhanced
by a factor ${\cal A}^4$ over the geometrical
cross-section,
\begin{equation}
\sigma_{inel} = {\cal A}^4 \; ({d\sigma \over d \Omega})_{geom}
\end{equation}
where the amplification factor ${\cal A}$ is defined by,
\begin{equation}
{\cal A} = {\psi (R) \over \psi_{free} (R) } \: .
\end{equation}
where R is the radius of the string, $R \sim
\eta^{-1}$. This method has been applied in
ref.~\cite{Perkins91} and~\cite{Michael}.

\subsection{The equations of motion}
\label{sec-motion}
The fermionic part of the lagrangian $L_F$ is given in
terms of 16 dimensional spinors as defined in Appendix
A. We shall consider only one family in this work, and
in particular the electron family. The fermionic
lagrangian for only one family,
\begin{equation}
L_F = L^{(e)}_F = \bar{\Psi}_{16} \gamma^\mu D_\mu \Psi_{16} + L_M + L_x
\end{equation}
where $L_M$ is the mass term and $L_x$ is the
lagrangian describing quark to lepton transitions
through the $X$, $X'$, $Y$, $Y'$ and $X_s$ gauge
bosons in SO(10) and given by
(equation~\ref{eq:lagi}). The covariant derivative is
given by $D_\mu = \partial_\mu - i e A_{\mu , str}
\tau_{str}$ where $A_{\mu , str}$ is the gauge field
forming the string and $\tau_{str}$ is the string
generator given by equation
(\ref{eq:tauall}). Therefore, since $\tau_{str}$ is
diagonal, there will be no mixing of fermions around
the string. The lagrangian $L_F$ will split in a sum
of eight terms, one for each fermion of the family. In
terms of 4-spinors, this is
\begin{equation}
L_F = \sum_{i = 1}^{8} L_f^i + L_x \label{eq:fer}
\end{equation}
where $L_f^i = i \bar{\psi}_L^i \gamma^\mu D_\mu^L
\psi_L^i + i \bar{\psi}_L^{c,i} \gamma^\mu D_\mu^{Lc}
\psi_L^{c,i} + L_m^i \label{eq:lagf}$, and i runs over
all fermions of the given familly. One can show that
$i \bar{\psi}_L^{c,i} \gamma^\mu D_\mu^L \psi_L^{c,i} = i
\bar{\psi}_R^{i} \gamma^\mu D_\mu^R \psi_R^{i}$ and
$\tau_{str}^{Lc \: i} = \tau_{str}^{R \: i}$. Finally,
$L_x$ is given by equation (\ref{eq:lagi}). It is easy
to generalise to more families.

 From equations (\ref{eq:fer}) and (\ref{eq:lagi}) we
derive the equations of motions for the fermionic
fields. We take the fermions to be massless inside and
outside the string core. This a relevant assumption
since our methods apply for energies above the
confinement scale. We consider the case of free quarks
scattering from the string and coupling with electrons
inside the string core. Outside the string core, the
fermions feel the presence of the string only by the
presence of the gauge field.We are interested in the
elastic cross sections for all fermions and in the
cross-section for these quark decaying into
electron. The fermionic lagrangian given by equations
(\ref{eq:fer}) and (\ref{eq:lagi}) becomes,
\begin{eqnarray}
L_F(e, {q}) &=& i \bar{e}_L \gamma^\mu D_\mu^{e,L} e_L + i
\bar{e}_R \gamma^\mu D_\mu^{e,R} e_R \nonumber \\
&&+  i \bar{q}_L \gamma^\mu D_\mu^{q,L} q_L + i \bar{q}_R
\gamma^\mu D_\mu^{q,R} q_R \nonumber \\
&&- {g G^\mu \over 2\sqrt{2}} \bar{q}_L \gamma_\mu e^+_L
- {g G^{' \mu} \over 2\sqrt{2}} \bar{q}_R \gamma_\mu
e^+_R + H.C.
\end{eqnarray}
giving the following equations of motion,
\begin{eqnarray}
i \gamma^\mu D^{e,L}_\mu e_L + {g G'_\mu \over 2\sqrt{2}}
\gamma^\mu q^c_L &=&0 \nonumber \\
i \gamma^\mu D^{e,R}_\mu e_R + {g G_\mu \over 2\sqrt{2}}
\gamma^\mu q^c_R&=&0 \nonumber \\
i \gamma^\mu D^{{q^c},L}_\mu q^c_L + {g G'_\mu \over 2
\sqrt{2}} \gamma^\mu e^-_L &=&0 \nonumber \\
i \gamma^\mu D^{{q^c},R}_\mu q^c_R+ {g G_\mu \over 2
\sqrt{2}} \gamma^\mu e^-_R &=&0 \label{eq:motion}
\end{eqnarray}
which are valid everywhere. The covariant derivatives
$D^{e,(L,R)}_\mu = \partial_\mu + ie A_{\mu , str}
\tau_{str}^{e,(L,R)}$ and $D^{{q^c},(L,R)}_\mu =
\partial_\mu + ie A_{\mu , str} \tau_{str}^{q,(L,R)}$.  We
have $\tau_{str}^{R, u} = \tau_{str}^{L, u} =
\tau_{str}^{L, e} = \tau_{str}^{L, d} = {1 \over 10}$
and $\tau_{str}^{R, e} = \tau_{str}^{R, d} = {-3 \over
10}$ together with $\tau_{str}^{Lc, \: i} =
\tau_{str}^{R, \: i}$ and $\tau_{str}^{L, \: i} =
\tau_{str}^{Rc, \: i}$. $G_\mu$ and $G'_\mu$ stand for
$X_\mu$, $X'_\mu$, $Y'_\mu$ or $Y'_\mu$, depending on
the chosen quark.

Since these equations involve quarks and lepton
mixing, we do not find independent solution for the
quark and lepton fields. However, we can solve these
equations taking linear combinations of the the quark
and lepton fields, ${q_L^c \pm e_L}$ and ${q_R^c \pm
e_R}$. In this case, the effective gauge fields are
\begin{equation}
e \, (A_{\mu , str} \tau_{str}^{f_L} \pm G_\mu)
\end{equation}
 and
\begin{equation}
e \, (A_{\mu , str} \tau_{str}^{f_R} \pm G'_\mu)
\end{equation}
 respectively.

In order to make the calculations easier, we use a
top-hat theta component for $G$ and $G'$ within the
string core, since Perkins et Al. \cite{Perkins91}
have shown that the physical results are insensitive
to the core model used for the gauge fields mediating
baryon violating processes.

\subsection{The External Solution}
\label{sec-ext}

Outside the string core, the gauge field of the string
$A_{\mu , str}$  has only, from
equations~\ref{eq:gauge} and~\ref{eq:tophat}, a non
vanishing component $A_\theta = {1 \over er}
\tau_{str}$, and the effective gauge fields $G$ and
$G'$ are set to zero. Therefore the equations of
motion (\ref{eq:motion}) for $r > R$ become,
\begin{eqnarray}
i \gamma^\mu D^{e,L}_\mu e_L  &=&0 \nonumber \\
i \gamma^\mu D^{e,R}_\mu e_R &=&0 \nonumber \\
i \gamma^\mu D^{u,L}_\mu q^c_L &=&0 \nonumber \\
i \gamma^\mu D^{u,R}_\mu q^c_R &=&0  \label{eq:ext}
\end{eqnarray}
where the covariant derivatives $D^{e,(L,R)}_\mu =
\partial_\mu + ie A_{\mu , str} \tau_{str}^{e,(L,R)}$ and
$D^{{q^c},(L,R)}_\mu = \partial_\mu + ie A_{\mu , str}
\tau_{str}^{q,(L,R)}$.

We take the usual Dirac representation $e_L =
(0,\xi_e)$ , $e_R = (\chi_e,0)$ , $q^c_L =
(0,\xi_{q^c})$ and $q^c_R = (\chi_{q^c},0)$ and the
mode decomposition for the spinors $\xi_{q^c}$,
$\xi_e$, $\chi_{q^c}$ and $\chi_e$,
\begin{eqnarray}
\chi_{(e, {q^c})}(r,\theta) &=& \sum_{n=-\infty}^{n=+\infty}
\left ( \begin{array}{rl}
	&\chi_{1\, (e, {q^c})}^n (r)\\
         i &\chi_{2\, (e, {q^c})}^n (r) \, e^{i\theta}
\end{array}
\right )
e^{in\theta} \nonumber\\
\xi_{(e, {q^c})}(r,\theta) &=& \sum_{n=-\infty}^{n=+\infty}
\left ( \begin{array}{rl}
	&\xi_{1\, (e, {q^c})}^n (r)\\
         i &\xi_{2 \, (e, {q^c})}^n (r) \, e^{i\theta}
\end{array}
\right )
e^{in\theta} \label{eq:mode} \; .
\end{eqnarray}
 From appendix~\ref{sec-extap} we see that the fields
$\xi_{1,(e, {q^c})}^n$, $\xi_{2,(e, {q^c})}^n$,
$\chi_{1,(e, {q^c})}^n$ and $\chi_{2,(e, {q^c})}^n$
satisfy Bessel equations of order $n -
\tau_{str}^{R\, (e, {q^c})}$,  $n + 1 -
\tau_{str}^{R\, (e, {q^c})}$, $n -  {\tau_{str}^{L \,
(e, {q^c})}}$ and $n -  \tau_{str}^{R\, (e, {q^c})}$
respectively. The external solution becomes,
\begin{equation}
	\left ( \begin{array}{l}
\xi_{(e, {q^c})} (r,\theta)\\
\chi_{(e, {q^c})} (r, \theta)
	\end{array}
	\right )
 = \sum_{n=-\infty}^{n=+\infty}
	\left ( \begin{array}{rlcrl}
&(v_n^{(e, {q^c})} J_{n - \tau_{str}^{R\, (e, {q^c})}}
(\omega r) &+& v_n^{(e, {q^c})'} J_{-(n - \tau_{str}^{R\,
(e, {q^c})})} (\omega r)) & e^{i n \theta} \nonumber\\
i&(v_n^{(e, {q^c})} J_{n + 1 - \tau_{str}^{R\, (e,
{q^c})}} (\omega r) &-& v_n^{(e, {q^c})'} J_{-(n + 1-
\tau_{str}^{R\, (e, {q^c})})} (\omega r)) & e^{i (n + 1)
\theta} \nonumber\\
&(w_n^{(e, {q^c})} J_{n - \tau_{str}^{L \, (e,
{q^c})}} (\omega r) &+& w_n^{(e, {q^c})'} J_{-(n -
\tau_{str}^{L \, (e, {q^c})})} (\omega r)) & e^{i n
\theta} \nonumber\\
i&(w_n^{(e, {q^c})} J_{n + 1 - \tau_{str}^{L \, (e,
{q^c})}} (\omega r) &-& w_n^{(e, {q^c})'} J_{-(n + 1 -
\tau_{str}^{L \, (e, {q^c})})} (\omega r)) & e^{i (n +
1)\theta}
       \end{array}
	\right ) \; .
\label{eq:external}
\end{equation}
Therefore, outside the string core, we have got
independent solutions for the quark and electron
fields.

\subsection{The Internal Solution}
\label{sec-int}

Inside the string core, the gauge field of the string,
$A_\mu$, is set to zero whereas $G_\theta$ and
$G'_\theta$ take the value $2\sqrt{2} A$ and $2
\sqrt{2} A'$ respectively. Therefore, the equations of
motion (\ref{eq:motion}) become,
\begin{eqnarray}
i \gamma^\mu \partial_\mu e_L + {g G'_\mu \over 2\sqrt{2}}
\gamma^\mu q^c_L &=&0 \nonumber \\
i \gamma^\mu \partial_\mu e_R + {g G_\mu \over 2\sqrt{2}}
\gamma^\mu q^c_R&=&0 \nonumber \\
i \gamma^\mu \partial_\mu q^c_L + {g G'_\mu \over 2\sqrt{2}}
\gamma^\mu e^-_L &=&0 \nonumber \\
i \gamma^\mu \partial_\mu q^c_R+ {g G_\mu \over 2\sqrt{2}}
\gamma^\mu e^-_R &=&0 \; . \label{eq:fint}
\end{eqnarray}
Since these equations of motions involve quark-leptons
mixings, there are no independent solutions for the
quarks and electron fields. However, we get solutions
for the fields $\rho^\pm$ and $\sigma^\pm$ which are linear
combinations of the quarks and electron fields,
\begin{equation}
\rho^\pm = \chi_{q^c} \pm \chi_e \label{eq:ro}
\end{equation}
and
\begin{equation}
\sigma^\pm = \xi_{q^c} \pm \xi_e \; . \label{eq:sigma}
\end{equation}

Using the mode decomposition (\ref{eq:mode}) for the
fields $\rho^\pm$ and $\sigma^\pm$, the internal solution
becomes,
\begin{equation}
\left ( \begin{array} {rl}
	&\rho_{n1}^\pm \, e^{in\theta}\\
	i & \rho_{n2}^\pm \, e^{i(n+1)\theta}\\
	&\sigma_{n1}^\pm \, e^{in\theta}\\
	i & \sigma_{n2}^\pm \, e^{i(n+1)\theta}
	\end{array}
	\right )
\end{equation}
where $\rho_{n1}^\pm$ and $\rho_{n2}^\pm$ and
$\sigma_{n1}^\pm$ and $\sigma_{n2}^\pm$ are the upper and
lower components of the fields $\rho^\pm$ and  $\sigma^\pm$
respectively. They are given in terms of
hyper-geometric functions. From
appendix~\ref{sec-intap} we get,
\begin{equation}
\rho_{n1}^\pm = (kr)^{|n|} e^{-ikr} \sum_{j =
0}^{n=+\infty} \alpha^\pm_j {(2ikr)^j \over j!}
\end{equation}
where $k^2 = w^2 - (e A)^2$, $e = {g \over 2
\sqrt{2}}$. $\alpha^\pm_{j+1} = {(a^\pm + j) \over (b
+ p)} \alpha^\pm_j$ with $a^\pm = {1\over 2} + |n| \pm
{e A (2n+1) \over 2ik}$ and $b = 1 +
2|n|$. $\rho_{n2}^\pm$ can be obtained using the coupled
equation (\ref{eq:int}.2) of
appendix~\ref{sec-intap}. We find,
\begin{equation}
\rho_{n2}^\pm = - {1\over w} (kr)^{|n|} e^{-ikr} \sum_{j
= 0}^{n=+\infty} \alpha^\pm_j {(2ikr)^j \over j!} \,
\,({|n| - n \over r} -ik + {j\over r} \pm e A) \; .
\end{equation}
We get similar hyper-geometric functions for the
fields $\sigma_{n1}^\pm$ and $\sigma_{n2}^\pm$.

\subsection{Matching at the String Core}
\label{sec-match}

 From now on, we will do calculations for the
right-handed fields, the calculations for the
left-handed ones being straight-forward. Once we have
our internal and external solutions, we match them at
the string core. We must take the same linear
combinations of the quark and lepton fields outside
and inside the core, and must have continuity of the
solutions at $r = R$. The continuity of the solutions
at $r=R$ implies,
\begin{eqnarray}
(\chi_{1,q}^n \pm \chi_{1,e}^n)^{out} &=& \rho_{n1}^{\pm \: in} \\
(\chi_{2,q}^n \pm \chi_{2,e}^n)^{out} &=& \rho_{n2}^{\pm \: in} \; .
\end{eqnarray}
Nevertheless, we will have discontinuity of the first derivatives,
\begin{eqnarray}
({d\over dr} \mp eA ) \, \rho_{n2}^{\pm \: in} &=&
({d\over dr} - {\tau_{str}^{R\, (e, {q^c})} \over R})
\, (\chi_{2,q}^n \pm \chi_{2,e}^n)^{out} \\
({d\over dr} \pm eA ) \, \rho_{n1}^{\pm \: in} &=&
({d\over dr} + {\tau_{str}^{R\, (e, {q^c})} \over R})
\, (\chi_{1,q}^n \pm \chi_{1,e}^n)^{out} \; .
\end{eqnarray}
These equations lead to a relation between the
coefficients of the Bessel functions for the external
solution, as derived in Appendix~\ref{sec-matchap},
\begin{equation}
{v_n^{q'} \pm v_n^{e'} \over v_n^{q} \pm v_n^{e}} = {
w \, \l_n^\pm J_{n + 1 - \tau_R}(wR) + J_{n -
\tau_R}(wR) \over w \, \l_n^\pm J_{-(n + 1 -
\tau_R)}(wR) + J_{-(n - \tau_R)}(wR)}
\label{eq:relation}
\end{equation}
where
\begin{equation}
\l_n^\pm = {\sum_{j = 0}^{n=+\infty} \alpha^\pm_j
{(2ikr)^j \over j!} \over \sum_{j = 0}^{n=+\infty}
\alpha^\pm_j {(2ikr)^j \over j!} ({|n| - n \over r}
-ik + {j\over r} \pm e A)} \; .\label{eq:lambda}
\end{equation}
The relations (\ref{eq:relation}) and (\ref{eq:lambda}) are the matching
conditions at $r = R$.

\subsection{The Scattering Amplitude}
\label{sec-ampl}

In order to caculate the scattering amplitude, we
match our solutions to an incoming plane wave plus an
outgoing scattered wave at infinity. However, since
the internal solution, and therefore the matching
conditions at $r = R$, are given in terms of linear
combinations of quarks and leptons, we consider
incoming waves of such linear combinations. Let
$f_n^\pm$ denote the scattering amplitude for the mode
n, $f_n^+$ if we consider the scattering of (quarks +
electrons) and $f_n^-$ if we consider the scattering
of (quarks - electrons). Then the matching conditions
at infinity are,
\begin{eqnarray}
\lefteqn{ (-i)^n
\left ( \begin{array}{l}
	 J_n \\
        i  J_{n+1} \, e^{i \theta}\\
\end{array}
\right ) +
{f_n^\pm e^{ikr} \over\sqrt{r}}
\left ( \begin{array}{rl}
& 1\\
i & e^{i \theta}\\
\end{array}
\right ) = } \nonumber \\
&& \left ( \begin{array}{rlcrl}
&(v_n^q \pm v_n^{e}) J_{n - \tau_R } &+& (v_n^{q'} \pm
v_n^{e'}) J_{-(n - \tau_R)}& \\
i & ((v_n^q \pm v_n^{e }) J_{n + 1 - \tau_R} & + &
(v_n^{q'} \pm v_n^{e'}) J_{-(n + 1 - \tau_R)}) & e^{i
\theta}\\
\end{array}
\right ) \; .
\end{eqnarray}
Using then the large r forms for the Bessel functions,
\begin{equation}
J_\mu (\omega r) =\sqrt{2 \over \pi \omega r} \cos(\omega r -
{\mu \pi \over 2} - {\pi \over 4}) \: , \nonumber
\end{equation}
and matching the coefficients of $e^{i\omega r}$ we find,
\begin{eqnarray}
\lefteqn{\sqrt{2 \pi \omega} f_n^\pm e^{i {\pi \over 4}} =} \nonumber \\
&&\left \{ \begin{array}{l}
e^{-i{n\pi}} (e^{i\tau_R \pi} -1) + (v_n^{q'} \pm v_n^{e'}) e^{i(n-\tau_R )
{\pi \over 2}} (1- e^{-2i (n-\tau_R ) \pi}) \\
 e^{i{n\pi}} (e^{i(n -\tau_R) \pi} - e^{-in\pi }) + ( v_n^q \pm v_n^{e\pm })
e^{-i(n-\tau_R ) {\pi \over 2}} (1- e^{2i (n-\tau_R ) \pi})
\end{array}
\right. \; . \label{eq:ampl}
\end{eqnarray}
Matching the coefficients $e^{-i \omega r}$, we get
relations between the Bessel functions coefficients,
\begin{equation}
(v_n^q \pm v_n^e) = (1 - (v_n^{q'} \pm v_n^{e'}) e^{-
i (n - \tau_R) {\pi \over 2}}) \: e^{- i (n - \tau_R)
{\pi \over 2}} \; . \label{eq:vnp}
\end{equation}
The relations (\ref{eq:ampl}), (\ref{eq:vnp}),
(\ref{eq:relation}) and (\ref{eq:lambda}) determine
the scattered wave.

\vspace{.75cm}
\section{The Elastic Cross-Section}
\label{sec-elast}
When there is no baryon number violating processes
inside the string core, when the gauge fields
mediating quark to lepton transitions are set to zero,
we have elastic scattering. In this case, the
scattering amplitude reduces to,
\begin{equation}
f_n^{elast} = {1 \over\sqrt{2 \pi \omega }} \, e^{-i {\pi \over 4}} \left \{
\begin{array}{ll}
 e^{- i n \pi } \, (e^{i \tau_R \pi } - 1) & n\geq 0 \\
e^{i n \pi} \, (e^{- i \tau_R \pi} - 1) & n \leq -1
\end{array}
\right. \; .
\end{equation}
The elastic cross-section per unit length is given by
\begin{equation}
\sigma_{elast} = | \sum_{n = -\infty}^{+ \infty}
f_n^{elast} e^{i n \theta} |^2 \; .
\end{equation}
Using the relations $\sum_{n = a}^{+ \infty } e^{i n
x} = {e^{iax} \over 1 - e^{ix}}$ and  $\sum_{n = -
\infty}^{b} e^{i n x} = {e^{i b x} \over 1 - e^{-
ix}}$, we find the elastic cross-section to be
\begin{equation}
\sigma_{elast} = {1 \over2 \pi \omega} {\sin^2{\tau_R \pi} \over \cos^2{\theta
\over 2}} \; .
\end{equation}
This is an Aharonov-Bohm cross-section, and $\tau_R$
is the flux in the core of the string.

Now, remember that $\tau_{str}^{Lc, u} =
\tau_{str}^{L, u} = \tau_{str}^{Lc, e} =
\tau_{str}^{L, d} = {1 \over 10}$ and $\tau_{str}^{L,
e} = \tau_{str}^{Lc, d} = {-3 \over 10}$ and
$\tau_{str}^{Lc, \: i} = \tau_{str}^{R, \: i}$ and
$\tau_{str}^{L, \: i} = \tau_{str}^{Rc, \: i}$. Hence,

\begin{equation}
\sigma_{elast}^{e_L} = \sigma_{elast}^{d_{R}}  >
\sigma_{elast}^{e_R} = \sigma_{elast}^{u_{R}} =
\sigma_{elast}^{d_L} = \sigma_{elast}^{u_{L}} \; .
\end{equation}
We therefore have a marked asymmetry between
fermions. We have got a marked asymmetry between left
and right handed electrons, left and right handed down
quarks or, since $\sigma_{elast}^{i_{Lc}} =
\sigma_{elast}^{i_{R}}$ and $\sigma_{elast}^{i_{Rc}} =
\sigma_{elast}^{i_{L}}$, between left handed particle and
antiparticle, respectively right handed, for the
electron and the down quark. But we have equal cross
sections for right handed particles and left handed
antiparticles for the electrons and the down quark,
and equal cross sections for both left handed and
right handed up quarks an anti-quarks. This is a
marked feature of grand unified theories. If cosmic
strings are found it may be possible to use this
asymmetry to identify the underlying gauge symmetry.

\section{The Inelastic Cross-Section}
\label{sec-inelast}
The gauge fields $X$, $X'$, $Y$ and $Y'$ are now
'switched on'. In this case we are calculating the
baryon number violating cross-section. If we consider
identical beams of incoming pure  $\rho^+$ and $\rho^-$,
recalling that $\rho^\pm = \chi_{q^c} \pm \chi_e$, this
will ensure that we will have an incoming beam of pure
quark. Therefore, the scattering amplitude for the
quark field is given by half the difference of $f_n^+$
and $f_n^-$, and the scattering amplitude for the
electron field is given by half the sum of $f_n^+$ and
$f_n^-$. From equation (\ref{eq:ampl}) we get,
\begin{equation}
\frac{1} {2} \,\sqrt{2 \pi \omega } \, (f_n^+ - f_n^- ) \, e^{i {\pi \over 4}}
= v_n^e \, e^{- i (n -\tau_R){\pi \over 2}} \, (1 - e^{- \tau_R 2 \pi}) \; .
\label{eq:fn}
\end{equation}
The inelastic cross-section for the quark field is given by,
\begin{equation}
\sigma_{inel} =  | \sum_{n = -\infty}^{+ \infty} (f_n^+ - f_n^-) \,  e^{i n
\theta} |^2 .
\end{equation}
Hence, from equation (\ref{eq:fn}),
\begin{equation}
\sigma_{inel} \sim  {1 \over \omega} | \sum_{n = -\infty}^{+ \infty} v_n^e \,
e^{- i n ({\pi \over 2} - \theta)} |^2 \; . \label{eq:inel}
\end{equation}
Using equations (\ref{eq:relation}), (\ref{eq:lambda}) and (\ref{eq:vnp}), we
find,
\begin{equation}
v_n^e = {e^{i(n - \tau_R) {\pi \over 2}} \over 2} ({1 \over \delta_n^+ + e^i (n
- \tau_R \pi)} - {1 \over \delta_n^- + e^i (n - \tau_R \pi)}) \label{eq:vninel}
\end{equation}
where
\begin{equation}
\delta_n^\pm= { w \, \l_n^\pm J_{n + 1 - \tau_R}(wR) + J_{n - \tau_R}(wR) \over
w \, \l_n^\pm J_{-(n + 1 - \tau_R)}(wR) + J_{-(n - \tau_R)}(wR)}
\label{eq:delta}
\end{equation}
and $\lambda^\pm$ are given by equations
(\ref{eq:lambda}). Equations (\ref{eq:inel}),
(\ref{eq:vninel}) and (\ref{eq:delta}) determine the
inelastic cross-section.This is given in terms of a
power series. However, using small argument expansions
for Bessel functions, we conclude that this power
series involves always one dominant term, the other
terms being suppressed by a factor $(\omega R)^n$ where n
is an integer such that $n \ge 1$. Therefore the
inelastic cross-section involves one dominant mode,
the other modes being exponentially suppressed. If $d$
denotes the dominant mode we get $\sigma_{inel} \sim
{1\over \omega} \, |v_{d}^e|^2$. The value of the dominant
mode depends on the sign of the the fractional flux
$\tau_{str}$. Our results can be summarised as
follow.

For $0 < \tau_R < 1$, the mode $n = 0$ is enhanced, and the other modes are
exponentially suppressed. Hence,
\begin{equation}
\sigma_{inel} \sim {1\over \omega} \, |v_0^e|^2 \; .
\end{equation}
Using small argument expansions for Bessel functions, this yields
\begin{equation}
\sigma_{inel} \sim {1\over \omega} \, (eA R)^2 \, (\omega R)^{4(1 - \tau_R)}
\end{equation}
where $A$ is the value of the gauge field inside the
string core, $e$ is the gauge coupling constant, and
$R \sim \eta$, $\eta$ being the the grand unified
scale$\sim 10^{15} GeV$. The greater amplification
occurs for $eA R \sim 1$, giving $\sigma_{inel} \sim
{1\over \omega} \, (\omega R)^{4(1 - \tau_R)}$.

For $-1 < \tau_R < 0$, the mode $n = -1$ is enhanced, and the other modes are
exponentially suppressed. Hence,
\begin{equation}
\sigma_{inel} \sim {1\over \omega} \, |v_{-1}^e|^2 \; .
\end{equation}
Using small argument expansions for Bessel functions, this yields
\begin{equation}
\sigma_{inel} \sim {1\over \omega} \, (eA R)^2 \, (\omega R)^{4 (1 + \tau_R)}
\; .
\end{equation}
The greater amplification occurs for $eA R \sim 1$,
giving $\sigma_{inel} \sim {1\over \omega} \, (\omega R)^{4(1 +
\tau_R)}$. Thus, the baryon number violating
cross-section is not a strong interaction
cross-section but is suppressed by a factor depending
on the grand unified scale $\eta \sim R^{-1} \sim
10^{15} GeV$. The baryon number violation
cross-sections are very small. For $u_L$ and $d_L$ we
obtain,
\begin{equation}
\sigma_{inel} \sim {1\over \omega} \, (\omega R)^{3.6} \; .
\end{equation}
Whereas for $d_R$ we get,
\begin{equation}
\sigma_{inel} \sim {1\over \omega} \, (\omega R)^{2.8} \; .
\end{equation}
Here again we have a marked asymmetry between left and
right handed fields. We find an indeterminate solution
for the left-conjugate up quark because its phase
around the string ($1 \over 10$) differs from the
phase of the left-handed electron $({-3\over 10})$ by
a fractional value different from a half.

\section{The second quantised cross-section}
\label{sec-second}

We now derive the baryon number violating
cross-sections using the perturbative method
introduced in section~\ref{sec-cs}.

Firstly, we calculate the geometrical
cross-section. This is the cross-section for free
fields $\psi_{free}$, where $\psi_{free}$ is a
2-spinor. In the case of gauge fields mediating
catalysis it is given by,
\begin{equation}
({d \sigma \over d \O })_{geom} = {1 \over \omega } \, (\omega R)^4 \, (eAR)^2
\end{equation}
where $\omega$ is the energy of the massless field
$\psi_{free}$, A is the value of the gauge field
mediating quark to lepton transitions, $e$ is the
gauge coupling constant and R is the radius of the
string with $R\sim \eta^{-1}$ with $\eta \sim 10^{15}$
GeV.

The second step is to calculate the amplification
factor $ {\cal A} = {\psi \over \psi_{free}}$, $\psi$
and $\psi_{free}$ being two 2-spinors. The catalysis
cross-section is enhanced by a factor ${\cal A}^4$
over the geometrical cross-section,
\begin{equation}
\sigma_{inel} \sim {\cal A}^4 \, ({d \sigma \over d \O})_{geom} \; .
\end{equation}
We now use the results of sections~\ref{sec-ext}, ~\ref{sec-int}
and~\ref{sec-match} where we have solved the equations of motion for the fields
$\psi$ and calculated the matching conditions. Using
equation~\ref{eq:external}, we get the wave function $\psi$ at the string core,
and for the mode n,
\begin{equation}
\psi^n  =
\left ( \begin{array}{lllll}
&((v_n^q \pm v_n^e) \, J_{n - \tau_{str}}(\omega R) &+& (v_n^{q'} \pm v_n^{e'})
\, J_{- (n  - \tau_{str})}(\omega R)) & e^{i n \theta}\\
i&((v_n^q \pm v_n^e) \, J_{n + 1- \tau_{str}}(\omega R) &+& (v_n^{q'} \pm
v_n^{e'}) \, J_{-(n + 1 - \tau_{str})}(\omega R)) & e^{i (n + 1) \theta}
\end{array}
\right )
\end{equation}
Using equations (\ref{eq:relation}) and
(\ref{eq:lambda}) and using small argument expansions
for Bessel functions, we conclude that for $n \ge 0$,
$(v_n^q \pm v_n^e) \gg (v_n^{q'} \pm v_n^{e'})$, and
for $n < 0$, $(v_n^q \pm v_n^e) << (v_n^{q'} \pm
v_n^{e'})$. Now, from equation (\ref{eq:vnp}), we see
that is one coefficient dominates that will be the
$O(1)$. Hence, for $n \ge 0$, $(v_n^q \pm v_n^e) \sim
1$, and for $n < 0$, $(v_n^{q'} \pm v_n^{e'}) \sim
1$. Therefore, using small argument expansions for
Bessel functions we get for $n \ge 0$,
\begin{equation}
\psi^n \sim
\left ( \begin{array}{l}
(\omega R)^{n - \tau_{str}} \\
(\omega R)^{n + 1 - \tau_{str}}
\end{array}
\right )
\end{equation}
which is to be compared with $\psi_2^{free} \sim 1$ for free spinors. The upper
component of the spinor is amplified while the other one is suppressed by a
factor $\sim (\omega R)$. For $n < 0$ we have,
\begin{equation}
\psi^n \sim
\left ( \begin{array}{l}
(\omega R)^{- (n - \tau_{str, R})} \\
(\omega R)^{- (n + 1 - \tau_{str, R})}
\end{array}
\right )
\end{equation}
Hence we conclude that for $n < 0$ the lower component is amplified while the
upper one is suppressed by a factor $\sim \omega R$.

Therefore, for $\tau_{str} = {- 3 \over 10}$, the amplification occurs for the
lower component and for the mode $n = -1$. The amplification factor is
\begin{equation}
{\cal A} \sim (\omega R)^{\tau_{str}}
\end{equation}
leading to the baryon number violating cross-section,
\begin{equation}
\sigma_{inel} \sim {1\over \omega} \, (eA R)^2 \, (\omega R)^{4 \, (1 +
\tau_{str})} \; .
\end{equation}
In the case $\tau_{str} = {1 \over 10}$, the
amplification occurs for the upper component and for
the mode $n = 0$. The amplification factor is,
\begin{equation}
{\cal A} \sim (\omega R)^{- \tau_{str}}
\end{equation}
leading to the baryon number violating cross-section,
\begin{equation}
\sigma_{inel} \sim {1\over \omega} \, (eA R)^2 \, (\omega R)^{4 \, (1 -
\tau_{str})} \; .
\end{equation}
This method shows explicitly which component of the
spinor and which mode are enhanced. The results agree
with scattering cross-sections derived using the first
quantised method.

\section{Conclusion}
\label{sec-concl}

We have investigated elastic and inelastic scattering
off abelian cosmic strings arising during the phase
transition $SO(10)
\stackrel{<\phi_{126}>}{\rightarrow} SU(5) \times Z_2$
induced by the Higgs in the 126 representation in the
early universe. The cross-sections were calculated
using both first quantised and second quantised
methods. The results of the two methods are in good
agreement.

During the phase transition $SO(10) \rightarrow SU(5)
\times Z_2$, only the right-handed neutrino gets a
mass. This together with the fact that we are
interested in energies above the confinement scales
allows us to consider massless particles.

The elastic cross-sections are found to be
Aharonov-Bohm type cross-sections. This is as
expected, since we are dealing with fractional
fluxes. We found a marked asymmetry between
left-handed and right-handed fields for the electron
and the down quark fields. But there is no asymmetry
for the up quark field. This is a general feature of
grand unified theories. If cosmic strings were
observed it might be possible to use Aharonov-Bohm
scattering to determine the underlying gauge group.

The inelastic cross-sections result from quark to
leptons transitions via gauge interactions in the core
of the string. The catalysis cross-sections are found
to be quite small, and here again we have a marked
asymmetry between left and right handed fields. They
are suppressed from a factor $\sim \eta^{-3.6}$ for
the left-handed up and down quark fields to a factor
$\sim \eta^{-2.8}$ for the right-handed down quark
field.

Previous calculations have used a toy model to
calculate the catalysis cross-section. Here the string
flux could be 'tuned' to give a strong interaction
cross-section. In our case the flux is given by the
gauge group, and is fixed for each particle
species. Hence, we find a strong sensitivity to the
grand unified scale. Our small cross-sections make it
less likely that grand unified cosmic strings could
erase a primordial baryon asymmetry, though they could
help generate it~\cite{Hindmarsh}.  If cosmic strings
are observed our scattering results, with the
distinctive features for the different particle
species, could help tie down the underlying gauge
group.

\section*{Acknowledgements}
We would like to thank T.W. Kibble, W. Perkins and
G.G. Ross for useful discussions, and PPARC for
financial support.

\appendix
\section{Brief review of SO(10)}
\label{sec-so10}

The fundamental representation of SO(10) consists of
10 generalised gamma matrices. They can be written in
an explicit notation, in terms of cross products,
\begin{eqnarray}
\Gamma_1 &=& \sigma_1 \times \sigma_3 \times \sigma_3 \times \sigma_3 \times
\sigma_3 \nonumber\\
\Gamma_2 &=& \sigma_2 \times \sigma_1 \times \sigma_3 \times \sigma_3 \times
\sigma_3 \nonumber\\
\Gamma_3 &=& I \times \sigma_1 \times \sigma_3 \times \sigma_3 \times \sigma_3
\nonumber\\
\Gamma_4 &=& I \times \sigma_2 \times \sigma_3 \times \sigma_3 \times \sigma_3
\nonumber\\
\Gamma_5 &=& I \times I \times \sigma_1 \times \sigma_3 \times \sigma_3
\nonumber\\
\Gamma_6 &=& I \times I \times \sigma_2 \times \sigma_3 \times \sigma_3
\nonumber\\
\Gamma_7 &=& I \times I \times I \times \sigma_1 \times \sigma_3 \nonumber\\
\Gamma_8 &=& I \times I \times I \times \sigma_2 \times \sigma_3 \nonumber\\
\Gamma_9 &=& I \times I \times I \times I \times \sigma_1 \nonumber\\
\Gamma_{10} &=& I \times I \times I \times I \times \sigma_2
\end{eqnarray}
where the $\sigma_i$ are the Pauli matrices and I denotes
the two dimensional identity matrix. They generate a
Clifford algebra defined by the anticommutation rules
\begin{equation}
\{ \Gamma_i , \Gamma_j \} = 2 \, \delta_{ij} \, \, \, \: \: i = 1,...,10 \; .
\end{equation}
One can define the chirality operator $\chi$, which is the generalised
$\gamma_5$ of the standard model by
\begin{equation}
\chi = (-i)^5 \prod_{i = 1}^{10} \Gamma_i \; .
\end{equation}
In terms of the cross-product notation, $\chi$ has the form,
\begin{equation}
\chi= \sigma_3 \times \sigma_3 \times \sigma_3 \times \sigma_3 \times \sigma_3
\; .
\end{equation}
The 45 generators of SO(10) are also given in terms of
the generalised gamma matrices
\begin{equation}
M_{ab} = {1 \over 2i} \, [\Gamma_i, \Gamma_j] \, \, \, i,j = 1... 10 \; .
\end{equation}
They are antisymmetric, purely imaginary $32 \times
32$ matrices. One can write the diagonal M,
\begin{eqnarray}
M_{12}&=& {1\over 2}\, \sigma_3 \times I \times I \times I
\times I \nonumber\\
M_{34}&=& {1\over 2}\, I \times \sigma_3 \times I \times I
\times I \nonumber\\
M_{56}&=& {1\over 2} \,I \times I \times \sigma_3 \times I
\times I \nonumber\\
M_{78}&=& {1\over 2}\, I \times I \times I \times \sigma_3
\times I \nonumber\\
M_{910}&=& {1\over 2}\, I \times I \times I \times I
\times \sigma_3  \; .
\end{eqnarray}
In SO(N) gauge theories fermions are conventionally
assigned to the spinor representation. For N even, the
spinor representation is $2^{N\over 2}$ dimensional
and decomposes into two equivalent spinors of
dimension $2^{{N\over 2} - 1}$ by means of the
projection operator $P = {1 \over 2} \, (1 \pm \chi)$,
where 1 is the $2^{N\over 2} \times  2^{N\over 2}$
identity matrix. Thus SO(10) has got two irreducible
representations,
\begin{equation}
\sigma^\pm = {1 \pm \chi \over 2}
\end{equation}
of dimension 16. Therefore SO(10) enables us to put
all the fermions of a given  family in the same
spinor. Indeed, since each family contains eight
fermions, we can put all left and right handed
particles of a given family in the same 16 dimensional
spinor. This is the smallest grand unified group which
can do so. However, gauge interactions conserve
chirality. Indeed,
\begin{equation}
\bar{\psi} \gamma_\mu A^\mu \psi = \bar{\psi_L} \gamma_\mu
A^\mu \psi_L + \bar{\psi_R} \gamma_\mu A^\mu \psi_R  \;
. \nonumber
\end{equation}
Therefore $\psi_L$ and $\psi_R$ cannot be put in the
same irreducible representation. Hence, instead of
choosing $\psi_L$ and $\psi_R$, we chose $\psi_L$ and
$\psi_L^c$. The fields $\psi_L$ and $\psi_L^c$
annihilate left-handed particles and antiparticles,
respectively, or create right-handed antiparticles and
particles. The fields $\psi_L$ and $\psi_L^c$ are
related to the fields $\psi_R$ and $\bar{\psi_R}$ by
the following relations,
\begin{eqnarray}
\psi_L^c &\equiv & P_L \psi^c = P_L C \bar{\psi}^T = C
(\bar{\psi} P_L )^T = C \bar{\psi_R}^T = C \gamma_0^T
\psi_R^* \\
\bar{\psi_L^c} &\equiv & \psi_L^{c \dagger} \gamma_0 =
\psi_R^{* \dagger} \gamma_0^{T \dagger} C^\dagger \gamma_0 = -
\psi_R^T C^{-1} = \psi_R^T C
\end{eqnarray}
where the projection operators $P_{L,R} = {\frac{1} {2}} (1
\pm \gamma_5)$ and C is the usual charge conjugation
matrix. For the electron family we get,
\begin{equation}
\Psi^{(e)}_L = (\nu_{(e)}^c\, , u^c_r\, , u^c_y\, ,
u^c_b\, , d_b\, , d_y\, , d_r\, , e^-\, , u_b\, ,
u_y\, , u_r\, , \nu_{(e)}\, , e^+\, , d^c_r\, ,
d^c_y\, , d^c_b)_L\\ \label{eq:psiL}
\end{equation}
where the upper index c means conjugate, and the
sub-indices refer to quark colour. We find similar
spinor $\Psi^{(\mu)}$ and $\Psi^{(\tau)}$ associated
with the $\mu$ and the $\tau$ family respectively:
\begin{eqnarray}
\Psi^{(\mu)} &=& (\nu_{(\mu)}^c\, , c^c_r\, , c^c_y\,
, c^c_b\, , s_b\, , s_y\, , s_r\, , \mu^-\, , c_b\, ,
c_y\, , c_r\, , \nu_{(\mu)}\,, \mu^+\, , s^c_r\, ,
s^c_y\, , s^c_b)_L\\
\Psi^{(\tau)} &=& (\nu_{(\tau)}^c\, , t^c_r\, ,
t^c_y\, , t^c_b\, , b_b\, , b_y\, , b_r\, , \tau^-\, ,
t_b\, , t_y\, , t_r\, , \nu_{(\tau)}\,, \tau^+\, ,
b^c_r\, , b^c_y\, , b^c_b)_L \; .
\end{eqnarray}

\section{The external solution}
\label{sec-extap}

We want to solve equations (\ref{eq:ext}). We set
$\partial_t = -i \omega$, where $\omega$ is the energy of the
electron and take the usual Dirac representation $e_L
= (0,\xi_e)$ , $e_R = (\chi_e,0)$ , $q^c_L =
(0,\xi_q)$ and $q^c_R = (\chi_q,0)$. We use the usual
mode decomposition for the spinors $\xi_q$, $\xi_e$,
$\chi_q$ and $\chi_e$ :
\begin{eqnarray}
\chi_{(e, {q^c})}(r,\theta) &=& \sum_{n=-\infty}^{n=+\infty}
\left ( \begin{array}{rl}
	&\chi_{1\, (e, {q^c})}^n (r)\\
         i &\chi_{2\, (e, {q^c})}^n (r) \, e^{i\theta}
\end{array}
\right )
e^{in\theta} \nonumber\\
\xi_{(e, {q^c})}(r,\theta) &=& \sum_{n=-\infty}^{n=+\infty}
\left ( \begin{array}{rl}
	&\xi_{1\, (e, {q^c})}^n (r)\\
         i &\xi_{2 \, (e, {q^c})}^n (r) \, e^{i\theta}
\end{array}
\right )
e^{in\theta} \; .
\end{eqnarray}
 Then, using the basis,
\begin{equation}
\gamma^j = \left ( \begin{array} {lr}
              0 & -i\sigma^j\\
	      i \sigma^j & 0
		\end{array}
	\right )
\end{equation}
the equations of motion (\ref{eq:ext}) become,
\begin{eqnarray}
\left. \begin{array}{cccccccccc}
\omega \chi_{1,(e, {q^c})}^n &-& ( {d\over dr} &+& { n+1
\over r} &-& {\tau_{str}^{R\, (e, {q^c})} \over r} )
&\chi_{2,(e, {q^c})}^n &=& 0 \\ [0.1cm]
\omega \chi_{2,(e, {q^c})}^n &+& ( {d\over dr}  &-& { n
\over r} &+& { \tau_{str}^{R \, (e, {q^c})} \over r} )
& \chi_{1,(e, {q^c})}^n &=& 0  \\ [0.1cm]
\omega \xi_{1,(e, {q^c})}^n &+& ( {d\over dr} &+& { n+1
\over r} &-& {\tau_{str}^{L \, (e, {q^c})} \over r} )
& \xi_{2,(e, {q^c})}^n &=& 0  \\ [0.1cm]
\omega \xi_{2,(e, {q^c})}^n &-& ( {d\over dr} &-& { n
\over r} &+& {\tau_{str}^{L \, (e, {q^c})}\over r} ) &
\xi_{1,(e, {q^c})}^n &=& 0 \label{eq:mot}
	\end{array}
	\right. \; .
\end{eqnarray}
It is easy to show that the fields $\xi_{1,(e,
{q^c})}^n$, $\xi_{1,(e, {q^c})}^n$, $\chi_{1,(e,
{q^c})}^n$ and $\chi_{2,(e, {q^c})}^n$ satisfy Bessel
equations of order $n -  \tau_{str}^{R\, (e, {q^c})}$,
$n + 1 -  \tau_{str}^{R\, (e, {q^c})}$, $n -
{\tau_{str}^{L \, (e, {q^c})}}$ and $n -
\tau_{str}^{R\, (e, {q^c})}$ respectively. Hence the
external solution is,
\begin{equation}
	\left ( \begin{array}{l}
\xi_{(e, {q^c})} (r,\theta)\\
\chi_{(e, {q^c})} (r, \theta)
	\end{array}
	\right )
 = \sum_{n=-\infty}^{n=+\infty}
	\left ( \begin{array}{rlcrl}
&(v_n^{(e, {q^c})} Z^1_{n - \tau_{str}^{R\, (e,
{q^c})}} (\omega r) &+& v_n^{(e, {q^c})'} Z^2_{n -
\tau_{str}^{R\, (e, {q^c})}} (\omega r)) & e^{i n \theta}
\nonumber\\
i&(v_n^{(e, {q^c})} Z^1_{n + 1 - \tau_{str}^{R\, (e,
{q^c})}} (\omega r) & +& v_n^{(e, {q^c})'} Z^2_{n + 1 -
\tau_{str}^{R\, (e, {q^c})}} (\omega r)) & e^{i (n + 1)
\theta} \nonumber\\
&( w_n^{(e, {q^c})} Z^1_{n - \tau_{str}^{L \, (e,
{q^c})}} (\omega r) & +& w_n^{(e, {q^c})'} Z^2_{n -
\tau_{str}^{L \, (e, {q^c})}} (\omega r)) & e^{i n \theta}
\nonumber\\
i &(w_n^{(e, {q^c})} Z^1_{n + 1 - \tau_{str}^{L \, (e,
{q^c})}} (\omega r) &+& w_n^{(e, {q^c})'} Z^2_{n + 1 -
\tau_{str}^{L \, (e, {q^c})}} (\omega r)) & e^{i (n + 1)
\theta}
       \end{array}
	\right ) \; .
\end{equation}
The order of the Bessel functions will always be
fractional. We therefore take $Z^1_\nu = J_\nu$ and
$Z^2_\nu = J_{-\nu}$.

\section{The internal solution}
\label{sec-intap}

We get solutions for fields which are linear
combinations of the quark and electron fields. Indeed,
we get solutions for the fields $\sigma^\pm = \xi_q \pm
\xi_e$ and $\rho^\pm = \chi_q \pm \chi_e$. Using the
mode decomposition (\ref{eq:mode}), the upper
components of the fields $\rho^\pm$ and  $\sigma^\pm$ are
respectively $\rho_{n1}^\pm =  \chi_{1\, {q^c}}^n \pm
\chi_{1\, e}^n$ and $\rho_{n2}^\pm =  \chi_{2\, {q^c}}^n
\pm \chi_{2\, e}^n$ whilst the lower components are
$\sigma_{n1}^\pm = \xi_{1\, {q^c}}^n \pm \xi_{1\, e}^n$
and $\sigma_{n2}^\pm =  \xi_{2\, {q^c}}^n \pm \xi_{2\,
e}^n$ respectively. The equations of motions
(\ref{eq:fint}) become
\begin{eqnarray}
\left.   \begin{array} {cccccccccc}
\omega \rho_{n1}^\pm & - &({d\over dr} & + & {n+1 \over r}
&\mp & e A')&\rho_{n2}^\pm &=&0 \\ [0.2cm]
\omega \rho_{n2}^\pm & + &({d\over dr} &-& {n \over r} &\pm
& e A' )&\rho_{n1}^\pm &=&0  \\ [0.2cm]
\omega \sigma_{n1}^\pm &+& ({d\over dr} &+& {n+1 \over r} &\mp
& e A )&\sigma_{n2}^\pm &=&0 \\ [0.2cm]
\omega \sigma_{n2}^\pm &-& ({d\over dr} &-& {n \over r} &\pm &
e A ) &\sigma_{n1}^\pm &=&0
\end{array}
	\right. \; . \label{eq:int}
\end{eqnarray}
Combining the two first equations of
 ({\ref{eq:int}), one can see that $\rho_{n1}^\pm$
satisfy an hyper-geometric equation giving,
\begin{equation}
\rho_{n1}^\pm = (kr)^{|n|} e^{-ikr} \sum_{j =
0}^{n=+\infty} \alpha^\pm_j {(2ikr)^j \over j!}
\end{equation}
where $k^2 = w^2 - (e A)^2$, $e = {g \over 2
\sqrt{2}}$. $\alpha^\pm_{j+1} = {(a^\pm + j) \over (b
+ p)} \alpha^\pm_j$ with $a^\pm = {1\over 2} + |n| \pm
{e A (2n+1) \over 2ik}$ and $b = 1 +
2|n|$. $\rho_{n2}^\pm$ can be obtained using the coupled
equation (\ref{eq:int}.2). We find
\begin{equation}
\rho_{n2}^\pm = - {1\over w} (kr)^{|n|} e^{-ikr} \sum_{j
= 0}^{n=+\infty} \alpha^\pm_j {(2ikr)^j \over j!} \,
\,({|n| - n \over r} -ik + {j\over r} \pm e A) \; .
\end{equation}
$\sigma_{n2}^\pm$ are also solutions of hyper-geometric
equations, and using the coupled equation
(\ref{eq:int}.4) we get,
\begin{eqnarray}
\sigma_{n1}^\pm &=& (kr)^{|n|} e^{-ikr} \sum_{j =
0}^{n=+\infty} \beta^\pm_j {(2ikr)^j \over j!} \\
\sigma_{n2}^\pm &=& - {1\over w} (kr)^{|n|} e^{-ikr}
\sum_{j = 0}^{n=+\infty} \beta^\pm_j {(2ikr)^j \over
j!} \, \,({|n| - n \over r} -ik + {j\over r} \pm e A')
\end{eqnarray}
where $k^2 = w^2 - (e A')^2$, $\beta^\pm_{j+1} =
{(c^\pm + j) \over (b + p)} \beta^\pm_j$ with $c^\pm =
{1\over 2} + |n| \pm {e A' (2n+1) \over 2ik}$.
And the internal solution is,
\begin{equation}
\left ( \begin{array} {rl}
	&\rho_{n1}^\pm \, e^{in\theta}\\
	i & \rho_{n2}^\pm \, e^{i(n+1)\theta}\\
	&\sigma_{n1}^\pm \, e^{in\theta}\\
	i & \sigma_{n2}^\pm \, e^{i(n+1)\theta}
	\end{array}
	\right ) \; .
\end{equation}
Therefore the internal solution is giving by a linear
combination of the quark and electron fields.

\section{The matching conditions}
\label{sec-matchap}

The continuity of the solutions at $r=R$ lead to,
\begin{eqnarray}
\lefteqn{(kR)^{|n|} \, e^{-ikR} \sum_{j =
0}^{n=+\infty} \alpha^\pm_j \, {(2ikR)^j \over j!}}
\label{eq:matchun} \nonumber \\
&& = \: (v_n^q \pm v_n^{e }) J_{n - \tau_R } (\omega R) \:
+ \: (v_n^{q'} \pm v_n^{e'}) J_{-(n - \tau_R)} (\omega R)
\end{eqnarray}
\begin{eqnarray}
\lefteqn{- {1\over w} \,  (kR)^{|n|} \, e^{-ikR}
\sum_{j = 0}^{n=+\infty} \alpha^\pm_j \, {(2ikr)^j
\over j!} \, \,({|n| - n \over R} -ik + {j\over R} \pm
e A)} \nonumber \\
&& = \: (v_n^q \pm v_n^{e }) J_{n + 1 - \tau_R} (\omega R)
\:  + \: (v_n^{q'} \pm v_n^{e'}) J_{-(n + 1 - \tau_R)}
(\omega R) \; . \label{eq:matchde}
\end{eqnarray}
Nevertherless, we will have discontinuity of the first
derivatives. Indeed, inside we have
\begin{eqnarray}
\left.   \begin{array} {cccccccccc}
\omega \rho_{n1}^\pm & - &({d\over dr} & + & {n+1 \over r}
&\mp & e A')&\rho_{n2}^\pm &=&0 \\
\omega \rho_{n2}^\pm & + &({d\over dr} &-& {n \over r} &\pm
& e A' )&\rho_{n1}^\pm &=&0
\end{array}
	\right.
\end{eqnarray}
whereas outside we have
\begin{eqnarray}
\left. \begin{array}{cccccccccc}
\omega (\chi_{1,{q^c}}^n \pm \chi_{1,e}^n) &-& ( {d\over
dr} &+& { n+1 \over r} &-& {\tau_{str}^{R\, (e,
{q^c})} \over r} ) &( \chi_{2,{q^c}}^n \pm
\chi_{2,e}^n) &=& 0 \\
\omega (\chi_{2,q}^n \pm \chi_{2,e}^n) &+& ( {d\over dr}
&-& { n \over r} &+& { \tau_{str}^{R\, (e, {q^c})}
\over r} ) & (\chi_{1,{q^c}}^n \pm \chi_{1,e}^n) &=& 0

	\end{array}
	\right. \; .
\end{eqnarray}
Now,
\begin{eqnarray}
(\chi_{1,{q^c}}^n \pm \chi_{1,e}^n)^{out} &=&
\rho_{n1}^{\pm \: in} \\
(\chi_{2,{q^c}}^n \pm \chi_{2,e}^n)^{out} &=&
\rho_{n2}^{\pm \: in}
\end{eqnarray}
giving us the relations for the first derivatives,
\begin{eqnarray}
({d\over dr} \mp eA ) \, \rho_{n2}^{\pm \: in} &=&
({d\over dr} - {\tau_{str}^{R\, (e, {q^c})} \over R})
\, (\chi_{2,{q^c}}^n \pm \chi_{2,e}^n)^{out}
\label{eq:matchdif}\\
({d\over dr} \pm eA ) \, \rho_{n1}^{\pm \: in} &=&
({d\over dr} + {\tau_{str}^{R\, (e, {q^c})} \over R})
\, (\chi_{1,{q^c}}^n \pm \chi_{1,e}^n)^{out} \; .
\end{eqnarray}
Dividing equation (\ref{eq:matchun}) by equation
(\ref{eq:matchde}) or either replacing equation
(\ref{eq:matchun}) in equation (\ref{eq:matchdif}), we
get the following relations
\begin{equation}
{v_n^{q'} \pm v_n^{e'} \over v_n^{q} \pm v_n^{e}} = {
w \, \l_n^\pm J_{n + 1 - \tau_R}(wR) + J_{n -
\tau_R}(wR) \over w \, \l_n^\pm J_{-(n + 1 -
\tau_R)}(wR) + J_{-(n - \tau_R)}(wR)}
\end{equation}
where
\begin{equation}
\l_n^\pm = {\sum_{j = 0}^{n=+\infty} \alpha^\pm_j
{(2ikr)^j \over j!} \over \sum_{j = 0}^{n=+\infty}
\alpha^\pm_j {(2ikr)^j \over j!} ({|n| - n \over r}
-ik + {j\over r} \pm e A)} \; .
\end{equation}


\begin{thebibliography} {99}

\bibitem{Shellard94} E.P.S. Shellard and A. Vilenkin
'Cosmic Strings and Other Topological Defects'
(Cambridge University Press, 1994)
\bibitem{Robert} Y.B. Zel'dovich MNRAS 192 (1980) 663;
A. Vilenkin Phys. Rev. Lett. {\bf 46}, 1169 (1981);
R. Brandenberger, Phys. Scripta {\bf T 36}, 114
(1991).
\bibitem{Joao} B. Allen, R.R. Caldwell,
E.P.S. Shellard, A. Stebbins and S. Veeraraghavan
FERMILAB-Conf-94/197-A (July 1994).
\bibitem{Rubakov} V. Rubakov JETP Lett. {\bf 33}
(1981), Nucl. Phys. {\bf 203}, 311 (1982); C. Callan
Phys. Rev. {\bf D 25}, 2141 (1982).
\bibitem{Branden88} R.H. Brandenberger, A.C. Davis and
A. Matheson Nucl. Phys. {\bf B 307}, 909 (1988).
\bibitem{Perkins91} W.B. Perkins, L. Perivolaropoulos,
A.C. Davis, R.H. Brandenberger and A. Matheson,
Nucl. Phys. {\bf B 353}, 237 (1991); M.G. Alford,
J. March-Russel and F. Wilczek, Nucl. Phys. {\bf B
328}, 140 (1989).
\bibitem{Bucher} M. Bucher and A. Goldhaber,
Phys. Rev. {\bf D49}, 4167 (1994).
\bibitem{Kib76} T.W. Kibble, Journal of Physics {\bf A
9}, 1387 (1976).
\bibitem{Warren} S. Dimopoulos, J. Preskill and
F. Wilczeck, Phys. Lett. {\bf B 119}, 320 (1982).
\bibitem{Branden89} R.H. Brandenberger, A.C. Davis and
A. Matheson, Phys. Lett. {\bf B 218}, 304 (1989);
A.Matheson. L. Perivolaropoulos, W. Perkins,
A.C. Davis and R.H. Brandenberger, Phys. Lett. {\bf B
248}, 263 (1990).
\bibitem{Alf89} M.G. Alford and F. Wilczek,
Phys. Rev. Lett. {\bf 62}, 1071 (1989).
\bibitem{Aharonov} Y. Aharonov and D. Bohm,
Phys. Rev. {\bf 115}, 485 (1959).
\bibitem{Adrian} T.W.B. Kibble Acta. Phys. Pol. {\bf B
13}, 723 (1982); R. Rohm, Ph.D. THESIS, Princeton
University (1985) unpublished.
\bibitem{G74} H. Georgi 'Particles and fields' (1974)
American Institute of Physics, New York.
\bibitem{Kibble82} T.W.B. Kibble, G. Lazarides and
Q.Shafi, Phys. Lett. {\bf B 113}, 237 (1982).
\bibitem{Hindmarsh} R. Brandenberger, A.C. Davis and
M. Hindmarsh, Phys. Lett. {\bf B 263}, 239 (1991).
\bibitem{Aryal87} M. Aryal and A. Everett,
Phys. Rev. {\bf D 35}, 3105 (1987); Chung-Pei Ma,
Phys. Rev. {\bf D 48}, 530 (1993).
\bibitem{barbieri80} R. Barbieri, D.V. Nanopoulos,
G. Morchio and F. Strocchi Phys. Lett. {\bf B 90} , 91
(1980).
\bibitem{Michael} M.A. Earnshaw and A.C. Davis,
Nucl. Phys. {\bf B 407}, 412 (1993).
\bibitem{Marie} M. Machacek, Nucl. Phys. {\bf B 159}, 37 (1979).
\end{thebibliography}
\end{document}